\documentclass[aip,reprint, jmp, amsmath, amssymb]{revtex4-1}
\usepackage{txfonts}
\usepackage{graphicx}
\usepackage{dcolumn}
\usepackage{amsmath}
\usepackage{subfigure}
\usepackage{bm}

\begin{document}

\title{Realization of spin-dependent splitting with arbitrary intensity patterns based on all-dielectric metasurfaces}

\author{Yougang Ke}
\author{Yachao Liu}
\author{Yongli He}
\author{Junxiao Zhou}
\author{Hailu Luo}
\email{hailuluo@hnu.edu.cn}
\author{Shuangchun Wen}
\affiliation{Laboratory for Spin Photonics, School of Physics and
Electronics, Hunan University, Changsha 410082, China}
\date{\today}

\begin{abstract}
We report the realization of spin-dependent splitting with arbitrary
intensity patterns based on all-dielectric metasurface. Compared to
the plasmonic metasurfaces, the all-dielectric metasurface exhibit
more high transmission efficiency and conversion efficiency, which
make it is possible to achieve the spin-dependent splitting with
arbitrary intensity patterns. Our findings suggest a way for
generation and manipulation of spin photons, and thereby offer the
possibility of developing spin-based nanophotonic applications.
\end{abstract}

\maketitle

The research field about optical metasurfaces, a class of optical
metamaterials with a reduced dimensionality, is rapidly expanding
recently, owing to their fascinating ability of controlling
light.~\cite{Shalaev2013} Most prominently, metasurfaces can create
an abrupt phase change over the scale of the subwavelength by
converting the incident polarization beam into corresponding
cross-polarization light.~\cite{Yu2011} By using the optical phase
discontinuities, a large number of applications have been proposed
and experimentally demonstrated such as
metasurfacelens,~\cite{Aieta2012,chen2012,Ni2013a} quarter-wave
metasurface plate~\cite{Zhao2011,Yu2012a}, generating vortex
beams~\cite{Karimi2014,Yang2014} and vector
beams,~\cite{Yi2014,Liu2014,Yi2015} metasurface
holograms,~\cite{Ni2013b,Zheng2015} to name just a few. In addition,
the planar geometry and ultra-thin nature of metasurfaces is
conducive to integrate into other nanodevices, and develop
miniaturized and ultra-compact optical devices. All these indicate
that metasurface is a prominent platform for designing ultra-thin
flat optical devices.

The high transmission efficiency of metasurface is a key factor for
the realization of practically metasurface-based optical devices.
The reflection-type metasurface based on the plasmon response of the
metallic nano-antenna, can achieve high transmission efficiency,
however, the transmission-type metasurface still have limits to
obtain high transmission efficiency, due to the ohmic losses in
metal.~\cite{Lin2014} All-dielectric metasurfaces without metallic
resonators have emerged as a new paradigm for introducing an abrupt
interfacial phase discontinuity.~\cite{Lin2014} Owing to their low
losses in the visible spectral range, all-dielectric metasurfaces
provide possibilities for the realization of practically high
quality transmission-type optical devices, such as forming clear
intensity patterns and realizing extremely low-loss beam splitting.
\begin{figure}
\includegraphics[width=8cm]{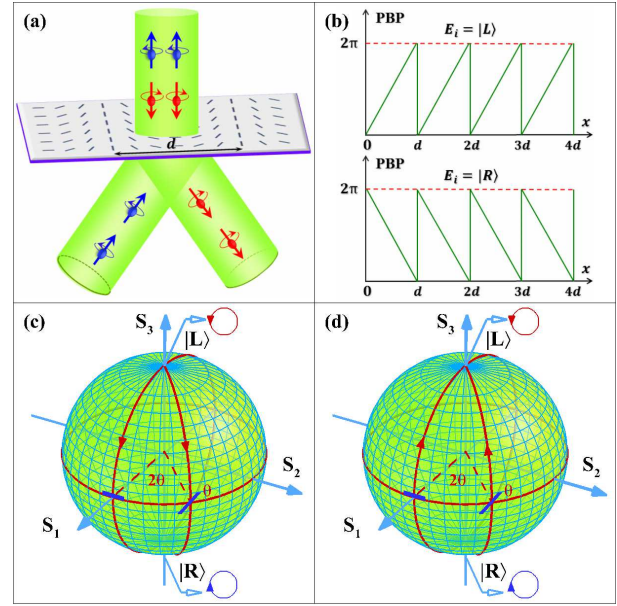}
\caption{\label{Fig1} (a) Schematic illustration of spin-dependent
splitting. The small balls with arrows represent the left- (red) and
right-handed (blue) photons, respectively. The metasurface reverses
the chirality of incident photons and steers normally incident
photons with opposite handedness to two directions, due to
space-variant PBP varying in one dimension. (b) The PBP for normally
incident $|L\rangle$ and $|R\rangle$ polarizations, respectively.
(c) and (d) show the visualization of the PBP on the Poincar\'{e}.
The longitude lines (red) with arrows represent the trajectories
corresponding the conversions of initial spin states to final states
in the metasurface with different directions of local optical axes
($0$ and $\theta$), and thereby the gained PBP is equal to half of
the area that is encompassed by the loop (red lines with arrows) on
the Poincar\'{e} sphere, and its absolute value can be represented
as $2\theta$ .}
\end{figure}

Polarizing beam splitters (PBSs) which can separate the two
orthogonal polarizations of light beam into different propagation
directions, are widely used in optical communications, imaging
systems, and optical recording. Conventional PBSs made by naturally
anisotropic materials require a large thickness to generate enough
walk-off distance between the two orthogonal polarizations owing to
the intrinsically small birefringence. Benefitting from the great
flexibility in engineering their optical properties, metamaterials
have been employed to construct polarization
splitter.~\cite{Luo2007,Zhao2008} Method based on transformation
optics which offers an unconventional approach to control
electromagnetic fields, have also been used to design splitter of
TE- and TM-polarized beams.~\cite{Kwon2008,Zhai2009} Despite great
progress in separating TE- and TM-polarized beams, it is still a
challenge to realize circular polarization beam splitter due to a
lack of natural materials with sufficient circular birefringence.
Recently, bulk metamaterials have been employed to block one handed
circular polarization beam while transmits the other handed
circularly polarized light.~\cite{Gansel2009,Turner2013} In
addition, based on gratings or metasurfaces, separation of
orthogonal circular polarizations appeared on the transmission side
of the elements have also been
demonstrated.~\cite{Bomzon2002,Hasman2002,Beresna2010,Huang2012,Lin2014,Ling2015}
However, almost all previous works about circular polarization
splitters did not consider the intensity pattern of separation,
which may have important applications in spin-based optical
information processing and imaging systems.

In this paper, we experimentally demonstrate spin splitting with
arbitrary intensity patterns, based on all-dielectric metasurface,
in combination with SLM. The light filed with desirable arbitrary
intensity profiles produced by SLM, is normally incident on the
metasurface. The metasurface with homogeneous phase retardation
$\pi$ and the orientation of local optical axis periodically varying
in one dimension creates the space-variant polarization state
manipulation, resulting in additional phase modifications. The
derived phase which is referred to PBP~\cite{Bomzon2002} creates the
spin-pendent phase gradient in one direction.  Metasurface steers
normally incident photons with opposite helicity to two directions
on the transmission side of the metasurface due to PBP gradient in
one dimension, as illustrated in Fig.~\ref{Fig1}.

It is convenient to character Pancharatnam-Berry phase optical
elements (PBOEs) using Jones calculus. By applying the optical
rotation matrix on the Jones matrix of a uniaxial crystal, one can
obtain a position-dependent Jones matrix for PBOEs, which can be
given by
\begin{equation}
\mathbf{T}(x,y)=\mathbf{R}(-\theta)\mathbf{J}\mathbf{R}(\theta)=\left(
  \begin{array}{cc}
    \cos(2\theta) & \sin(2\theta) \\
    \sin(2\theta) & -\cos(2\theta) \\
  \end{array}
\right),
\end{equation}
Where $\theta$ is the space-variant optical axis orientation of
PBOEs with homogeneous phase retardation $\pi$ and
$\mathbf{R}(\theta)$ the two-dimensional rotation matrix by angle
$\theta$.

When a left-circular ($|L\rangle$) and right-circular ($|R\rangle$)
polarization beam normally impinge onto the PBOEs, the output states
can be calculated as
\begin{eqnarray}
|E_{out}\rangle &=&\mathbf{T}(x,y)|L\rangle=\exp(i2\theta)|R\rangle, \\
|E_{out}\rangle
&=&\mathbf{T}(x,y)|R\rangle=\exp(-i2\theta)|L\rangle,
\end{eqnarray}
where $|L\rangle=(1,i)^{T}/\sqrt{2}$ and
$|R\rangle=(1,-i)^{T}/\sqrt{2}$ indicate the right- and
left-circular polarizations, respectively. Above equations show
unambiguously that PBOEs with constant phase retardance $\pi$ invert
the handedness of normally incident photons and introduce an
additional spin- and position-dependent phase $\pm2\theta$. The
additional space-variant phase is the so-called PBP. The PBP can be
conveniently visualized on the Poincar\'e sphere, as shown in
Figs.~\ref{Fig1}(c) and ~\ref{Fig1}(d). The two poles on the
Poincar\'e sphere represent the $|L\rangle$ and $|R\rangle$
polarization states, points on the equator indicate the linear
polarization states. When the $|L\rangle$ and $|R\rangle$
polarization beam normally impinge onto the metasurface with
constant phase retardance $\pi$, the beam at different points
evolute alone different longitude lines on the surface of the
Poincar\'e sphere to cross-polarization states, and then acquire a
phase modulation (PBP), due to space-varying optical axes in the
metasurface. The PBB is not a result of optical path differences,
but solely due to local changes in polarization.~\cite{Bomzon2002}
In addition, the additional phase depends on the spin sates of
incident photons and the orientation of local optical axes, which
suggests a route to manipulate the spin photons and enable many
spin-based photonics applications. Here, we focus on spin splitting.
\begin{figure}
\includegraphics[width=8cm]{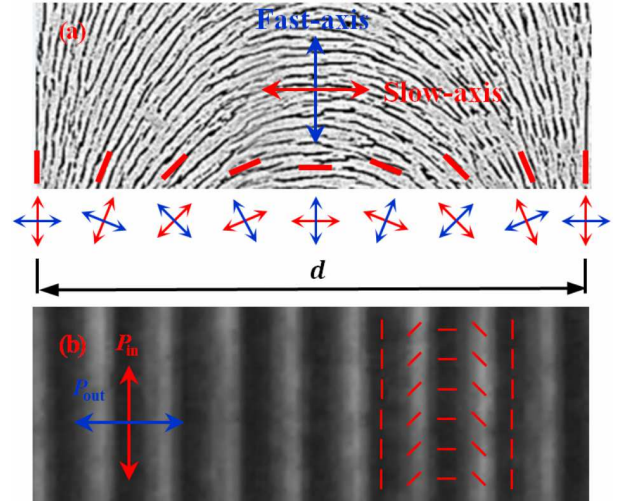}
\caption{\label{Fig2} (a) Scanning electron microscopy (SEM) image
of metasurface ($d=10$ $\mu$m). The cross with arrows below the SEM
image denote the local optical axis (fast-axis and slow-axis)
orientation in the place marked by red dashed lines. (b)
Cross-polarized optical images of optical-axis spatial distribution
in a structured metasurface applied in our experiments under crossed
linear polarizers. $P_{in}$ and $P_{out}$ denote the input and
output polarization states of light. Red dashed lines denote the
local optical axis direction.}
\end{figure}

To achieve this aim, we engineer the metasurface with homogeneous
phase retardation $\pi$ and the orientation of local optical axis
periodically varying in one dimension ($x$ direction) as shown in
Fig.~\ref{Fig1}(a). When $|L\rangle$ and $|R\rangle$ polarization
beam normally passing through the metasurface, the metasurface
introduces spacial variant PBP resulting in opposite linear phase
gradient in $x$ direction for incident $|L\rangle$ and $|R\rangle$
polarizations [Fig.~\ref{Fig1}(b)]. The constructed geometric phase
gradient ($\nabla_{x} \Phi$) steer different handedness photons to
opposite directions, which can be described by the following
expression:
\begin{equation}
\triangle k_{x}=\nabla_{x} \Phi=\frac{d \Phi}{d
x}=\frac{2\sigma_{\pm}d\theta}{dx}=2\sigma_{\pm} \Omega,
\end{equation}
where $\Delta k_{x}$ represents the shift occurs in $k$ space,
$\sigma_{\pm}=\pm 1$ is the incident spin state, $\Omega$ denotes
the spatial rotation of local optical axes in a unit length. The
shift in momentum-space will induce real-space shift upon
transmission. According to the mapping relationship between the
momentum space and real space, we get
\begin{equation}
\triangle x=\frac{2\sigma_{\pm} \Omega}{k_{0}}z,
\end{equation}
where  $\triangle x$ is the spin displacement, $k_{0}=2\pi/\lambda$
the free-space wave number and $z$ the transmission distance. Note
that the induced spin-dependent separation distance in real-space
increases linearly upon beam propagation and is equal to $2
|\triangle x|$.

The sample was designed to exhibit a uniform birefringent phase
retardation $\pi$ over the whole structured area at a wavelength of
$632.8$ nm. The distribution of its optical axis is shown to satisfy
$\theta=\Omega x$, where $\Omega=\pi/d$ is the spatial rotational
rate of the optical axis with the period $d=20$ $\mu$m. The sample
was prepared using femtosecond laser imprinting of space-variant
self-assembled form birefringence in silica glass
slab.~\cite{Shimotsuma2003,Beresna2011a,Beresna2011b}. The glass
substrate has a diameter of $25.4$ mm and the structured area
corresponded to a  $8$ mm $\times$ $8$ mm region centered on the
substrate (Altechna R$\&$D). Scanning electron microscope (SEM)
image for fabricated sample is the prominent way to character the
nanostructure. Despite its straightforwardness, the method limits
characterization of the sample created by femtosecond laser
nanostructuring of grass. To obtain the SEM image, it often requires
precise grinding and polishing of the glass
sample,~\cite{Beresna2011b} which frequently is followed with
chemical etching. This additional post-processing may destruct the
sample. Here, we provide SEM image of metasurface with the period
$d=10$ $\mu$m [see Fig.~\ref{Fig2}(a)]. An alternative,
nondestructive method to character the space-variant birefringence
structures embedded in silica glass is cross-polarized optical
images.~\cite{Hakobyan2014} The cross-polarized image of the sample
used in our experience as shown in Fig.~\ref{Fig2}(b), indicates the
optical axis direction periodically varying in one dimension. The
sample has a high transmission efficiency of $50.1\%$ and a high
conversion efficiency of $96.3\%$ (note that the transmission losses
are not taken into account when evaluating the conversion efficiency
from the measured data) at $632.8$ nm wavelength, measured by a
laser power meter.
\begin{figure}
\includegraphics[width=8.5cm]{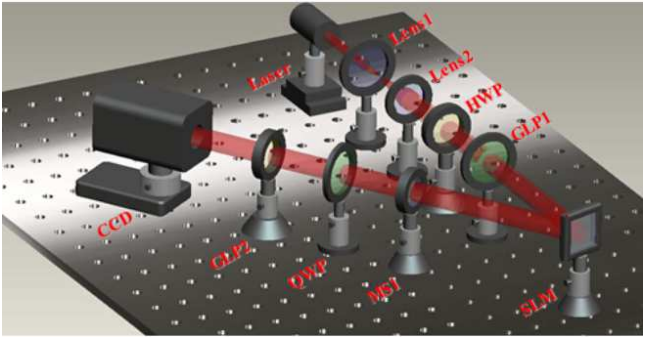}
\caption{\label{Fig3} Experiment setup for realizing spin-dependent
splitting with arbitrary intensity patterns. The light source is a
$21$ mW linearly polarized He-Ne laser at $632.8$ nm (Thorlabs
HNL210L-EC). Lens1 and Lens2 with effective focal length $25$ mm and
$75$ mm, respectively; HWP, half-wave plate (for adjusting the light
intensity); GLP1, Glan laser polarizer (for producing the
horizontally polarized light beam); SLM, phase-only spatial light
modulator (Holoeye PLUTO). The combination of a quarter-wave plate
(QWP), another Glan laser polarizer (GLP2), and a charge-coupled
device (CCD) is a typical setup to measure the Stokes parameter
$S_{3}$.}
\end{figure}

To demonstrate the spin-dependent splitting with arbitrary intensity
patterns, the experimental setup is depicted in Fig.~\ref{Fig3}. A
fundamental-mode Gauss beam with the diameter of beam waist $0.7$ mm
generated by He-Ne laser, first passes through a combination of a
short-focal-length lens (Lens1, focal length $25$ cm) and a
long-focal-length lens (Lens2, focal length $75$ cm), which is
collimated and expanded (the diameter of the beam waist: $2.1$ mm).
Then the light beam normally impinges onto half-wave plate (HWP)
which is used to control the light intensity to prevent the
charge-coupled devices (CCD) from saturation. The Glan laser
polarizer (GLP1) ensures the horizontally polarized light beam to
impinge into a phase-only spatial light modulator (SLM). Loading the
rationally designed phase picture into SLM (Holoeye PLUTO), we can
acquire desired patterns in light beam reflected from SLM. The
followed metasurface with local optical axis direction periodically
varying in one dimension splits the normally incident photons with
opposite handedness into two directions forming two identical
intensity patterns. Finally, the combination of a quarter-wave plate
(QWP), another Glan laser polarizer (GLP2), and a charge-coupled
device (CCD) is a typical setup to measure the Stokes parameter
$S_{3}$ of the separated light spots. The Stokes parameter $S_{3}$
[$S_{3}=(I_{\sigma+}-I_{\sigma-})/(I_{\sigma+}+I_{\sigma-})$, where
$I_{\sigma+}$ and $I_{\sigma-}$ represent the intensity of
 $|L\rangle$ and $|R\rangle$ polarization components, respectively.] can
be used to character the degree of the circular polarization with
$S_{3}=\pm 1$ corresponding to $|L\rangle$ and $|R\rangle$
polarization, respectively.~\cite{Born1999}
\begin{figure}
 \centering
 \subfigure{\includegraphics[width=2.4cm]{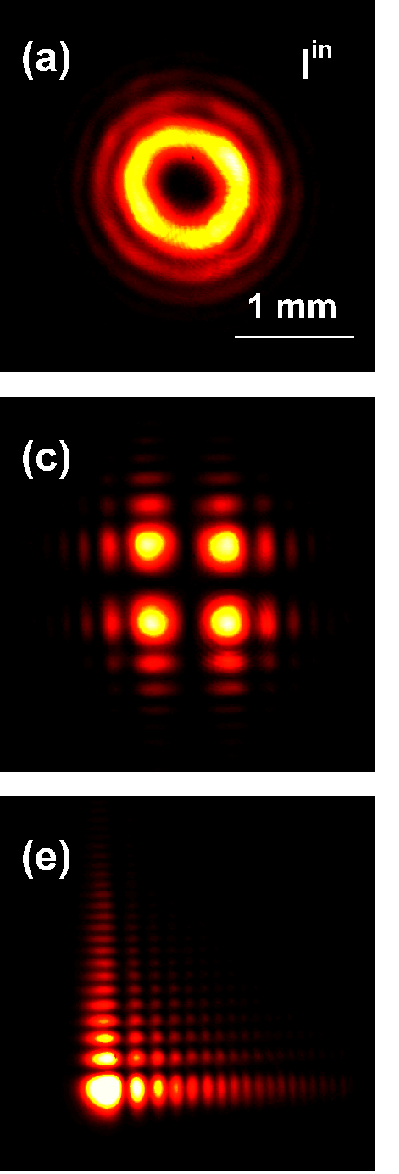}}
 \subfigure{\includegraphics[width=5.715cm]{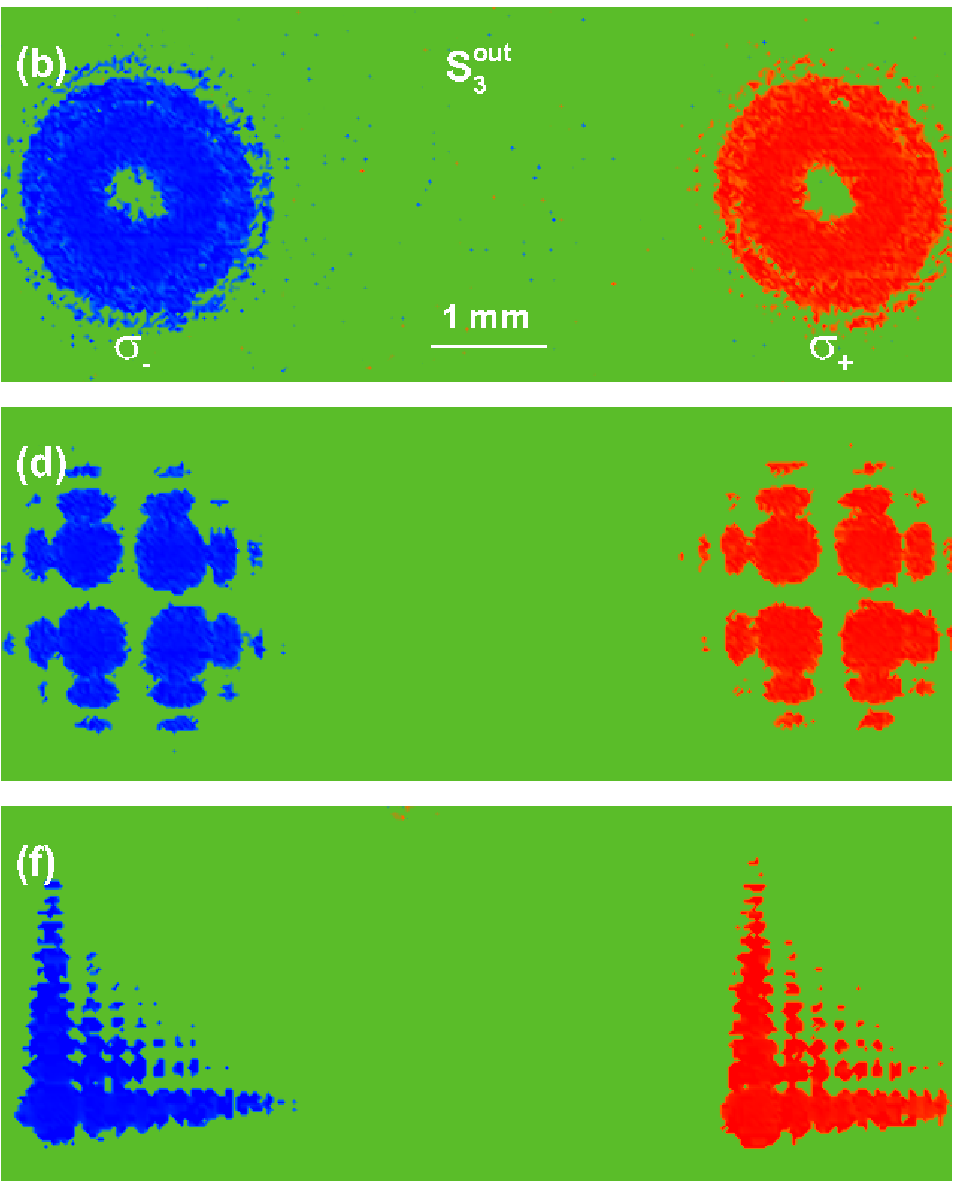}}
 \caption{ Intensity patterns (left column) of three typical linear
 polarization beams (vortex beam with topological charge $2$, Hermite-Gaussian
 beams, Airy beam in the order from top to bottom) before metasurface, and
 the corresponding normalized Stokes parameter $S_{3}$ (right column) of
 two spin-dependent splitting light spots after metasusrface.} \label{Fig4}
 \end{figure}

\begin{figure}
 \centering
 \subfigure{\includegraphics[width=2.4cm]{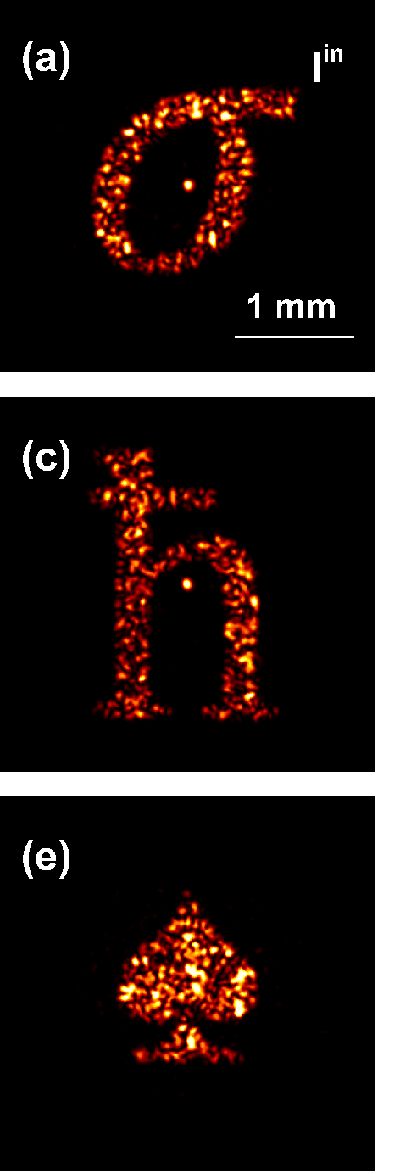}}
 \subfigure{\includegraphics[width=5.715cm]{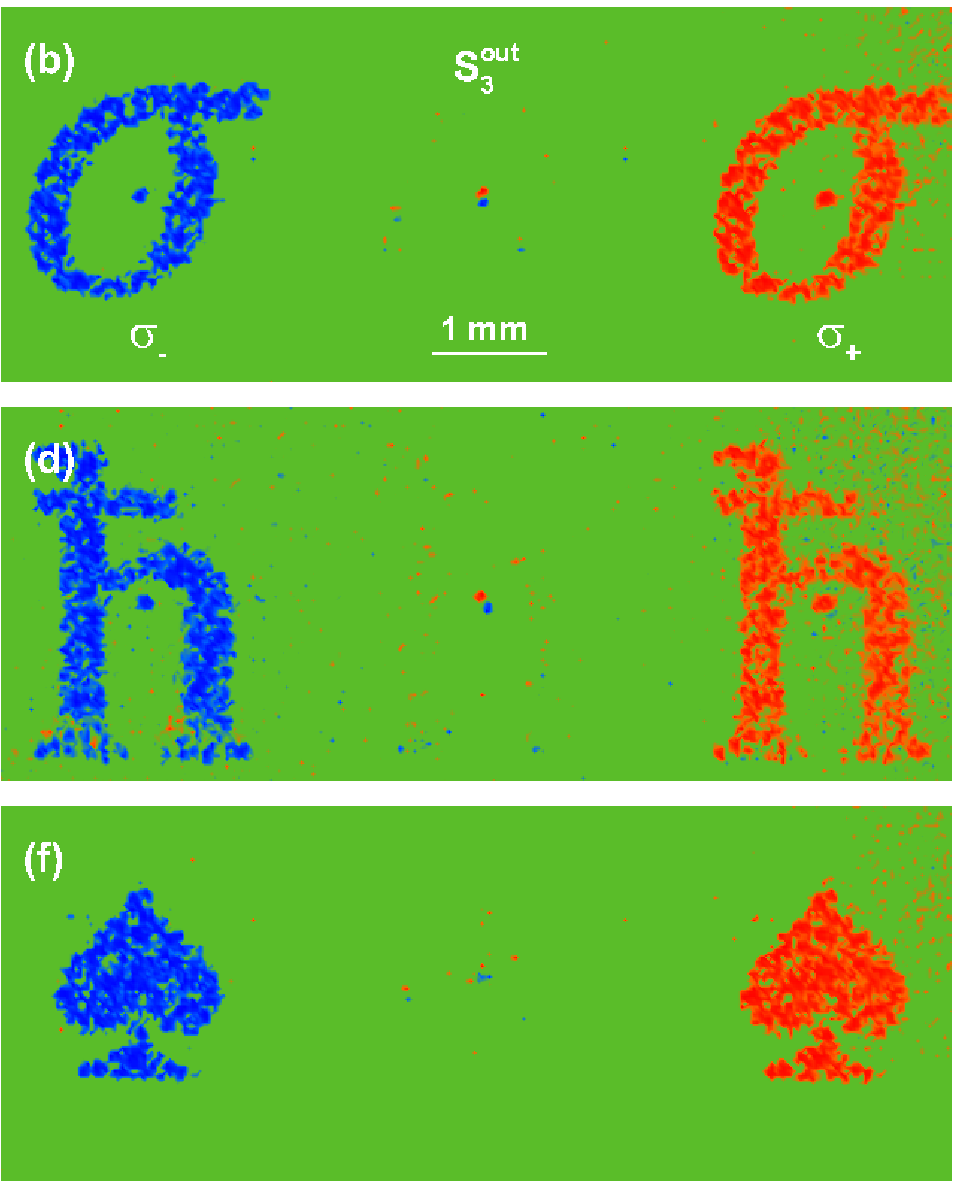}}
 \caption{ Intensity patterns (left column) of three typical linear
 polarization beams (two spin-dependent letters $\sigma$ and $\hbar$
and one $\spadesuit$ in the order from top to bottom) before
metasurface. The corresponding spin-dependent splitting spots after
metasurface is discriminated by normalized Stokes parameter $S_{3}$
(right column).} \label{Fig5}
 \end{figure}

To demonstrate the spin splitting with arbitrary intensity patterns,
we first use SLM to produce three important linear polarization
beams: vortex beam with a topological charge of $2$ (a doughnut
shape intensity distribution), Hermite-Gaussian beams (four lobes
with equal magnitudes), Airy beam (two main lobes and large side
lobes) as examples [Fig.~\ref{Fig4} (left column)]. The metasurface
creates opposite PBP gradient for normally incident left-handedness
and right-handedness photons [see Fig.~\ref{Fig1}(b)], respectively,
and defects the normally incident linear polarization beams to
opposite direction [Fig.~\ref{Fig1}(a)], resulting in two
spin-dependent spots with the same intensity pattern, due to
linearly polarized light can be viewed as equal-weight superposition
of the two spin state photons. It is convenient to character the
spin states of the two separated light spots using normalized Stokes
parameter $S_{3}$ [Fig.~\ref{Fig4} (right column)], because the
Stokes parameter $S_{3}$ can be used to describe the degree of
circular polarization. The profiles of the measured $S_{3}$
parameter [Fig.~\ref{Fig4} (right column)] keep typical character of
the intensity patterns of incident beams, which indicate obviously
the high quality of the generated patterns in the two spin-dependent
spots. The patterns of separated spots are not limited to typical
beam profiles, and we can achieve arbitrary patterns by loading the
rationally designed phase picture into SLM. Here, we produce two
spin-dependent letters ($\sigma$ and $\hbar$) and one $\spadesuit$
as further examples [Fig.~\ref{Fig5} (left column)], and the
resulting splitting patterns exhibit good performance as shown in
right column of Fig.~\ref{Fig5}.

In conclusion, we have proposed and experimentally demonstrated a
spin-dependent splitting with arbitrary intensity pattern,
consisting of a metasurface and a SLM. The metasurface has a high
transmission efficiency and a high conversion efficiency. The spin
splitting involving two types of phase: the dynamic phase and the
geometric PBP, which suggests combining the dynamic phase and PBP
may provide an additional degree of freedom to realize complex
spin-based photonics device. In addition, the spin splitter with
arbitrary intensity pattern may be used as spin analyzer and may
find potential in encoding information and spin encryption.

This research was partially supported by the National Natural
Science Foundation of China (Grants Nos. 11274106 and 11474089).

\end{document}